\documentclass[12pt,preprint]{aastex}
\usepackage{natbib,makeidx}

\newcommand{\swift}{Swift\,J1753.5-0127}

\slugcomment{Not to appear in Nonlearned J., 45.}

\shorttitle{The orbital period of Swift J1753.5-0127}
\shortauthors{Zurita et al.}

\begin{document}


\title{Swift  J1753.5-0127:  The Black Hole Candidate with  the shortest
orbital period}


\author{C. Zurita}
\affil{Instituto de Astronom\'{\i}a, UNAM, Mexico}
\email{czurita@iac.es}

\author{M. Durant}
\affil{Instituto de Astrof\'{\i}sica de Canarias, 38200 La Laguna, Tenerife, Spain}

\author{M.A.P. Torres}
\affil{Harvard-Smithsonian Center for Astrophysics, 60 Garden St, Cambridge, MA 02138, USA}

\author{T. Shahbaz}
\affil{Instituto de Astrof\'{\i}sica de Canarias, 38200 La Laguna, Tenerife, Spain}

\author{J. Casares}
\affil{Instituto de Astrof\'{\i}sica de Canarias, 38200 La Laguna, Tenerife, Spain}

\author{D. Steeghs}
\affil{Harvard-Smithsonian Center for Astrophysics, 60 Garden St, Cambridge, MA 02138, USA}
\affil{Department of Physics, University of Warwick, Coventry CV4 7AL, UK}


\begin{abstract}

\end{abstract}
We present time-resolved photometry  of the optical counterpart to the
black hole candidate \swift\, which has remained in the low/hard X-ray
state  and bright  at optical/IR  wavelengths since  its  discovery in
2005. At the  time of our observations  \swift\ does not  show a decay
trend  but  remains  stable  at   $R=16.45$  with  a  night  to  night
variability of $\sim$0.05\,mag.  The $R$-band light curves, taken from
2007 June  3 to August 31,  are not sinusoidal, but  exhibit a complex
morphology with  remarkable changes in shape and  amplitude.  The best
period  determination is  3.2443$\pm$0.0010  hours.  This  photometric
period is likely a superhump  period, slightly larger than the orbital
period.   Therefore, \swift\  is  the black  hole  candidate with  the
shortest  orbital period  observed  to date.   Our  estimation of  the
distance  is  comparable to  values  previously  published and  likely
places \swift\ in the Galactic halo.



\keywords{accretion, accretion disks ---
binaries: close --- stars:individual (Swift J1753.5-0127)--- X-rays: stars}

\section{Introduction}

X-ray  transients (XRTs)  are low-mass  accreting X-ray  binaries that
display  episodic X-ray  and  optical outbursts  whose luminosity  may
increase   by    several   orders   of    magnitude   from   quiescent
luminosity. Most  of them are believed  to harbor a black  hole as the
compact object.   Because of large changes in  the effective accretion
rate during the outburst, XRTs pass through certain identifiable X-ray
spectral states.

XRTs at maximum brightness are often  observed in a state in which the
X-ray emission  is dominated  by thermal emission  from the  hot inner
accretion disc  (i.e., the {\em high/soft} or  {\em thermal dominated}
state). Once the mass accretion rate drops below a critical value, the
spectrum switches to a state dominated by a hard non-thermal power law
component  (i.e., the  {\em low/hard}  state --  LHS).  At  ever lower
accretion rates  XRTs are in quiescence,  which may appear  as just an
extreme  example of  the LHS.   There  are, however,  systems that  in
outburst  do not  reach the  thermal dominated  state but  are instead
dominated  by  a   hard  non-thermal  power-law  component  \citep[see
e.g.,][]{brocksopp04}. The  persistance of the LHS  during an outburst
can be explained  by a truncation of the accretion  disc at some large
inner radius.   The interior  volume is filled  with a  hot, optically
thin,  quasi-spherical  accretion  flow,  where most  of  the  energy,
released  via viscous  dissipation, remains  in the  flow  rather than
being  radiated away  and is  finally advected  by the  compact object
\citep[e.g.,][]{narayan96}.   The  LHS  of  X-ray  binaries  has  been
associated with  jet activity \citep{fender01}. In the  context of the
coronal evaporation model \citep{meyer00} it is expected that the systems
with the  shortest orbital  periods remain in  the LHS  throughout the
outburst   \citep{meyerhofmeister04}.   Examples   of  such   sources  are
XTE\,J1118+480 \citep{hynes00}, in the Galactic Halo, and GRO J0422+32
\citep{sunyaev93}.\\

\swift\  was discovered by  the Swift Burst Alert  Telescope on
2005 Jun 30\footnote{The earliest detection of \swift\  in Palmer et
al. 2005  should be 2005-06-30 instead of  2005-05-30 (Palmer, private
communication)}  \citep{palmer05} as  a bright,  variable $\gamma$-ray
and hard X-ray transient source.   It was also detected in soft X-rays
with the  Swift X-ray  telescope (XRT) and  in UV with  the UV/Optical
Telescope  \citep{morris05,  still05}.   The optical  counterpart  was
found with  the MDM  2.4\,m telescope with  R$\sim$15.8 which  was not
visible  in  the  Digitized  Sky  Survey  \citep{halpern05} nor in the 
USNO B1.0 catalog ($V>$21).   Optical
spectroscopy  of \swift\ revealed  double-peaked H$\alpha$  and 
He{\sc ii} emission lines (Torres  et al.  2005a,b).  Furthermore, the source
was  detected in  radio likely  indicating jet  activity,  which would
usually be associated with a  LHS. However \swift\ does not follow the
usual  radio/X-ray  correlation  of  X-rays  binaries  in  this  state
\citep{cadolle07}.  INTEGRAL observations  of the source indicate that
the   spectrum  to   be  typical   of  a  Black Hole Candidate (BHC)  in   the   hard  state
\citep{cadolle07,  ramadevi07}.  Also, its power density spectrum shows a 0.6\,Hz quasi periodic oscillation with a shape tipically seen in BHCs \citep{morgan05}.  Analysis  of  the light  curves  and
hardness ratios of  \swift\ from the RXTE All-Sky  Monitor and pointed
RXTE observations reveal that the source has remained in the LHS since
the onset  of its  outburst in 2005.   However, it has  been suggested
from fits  to the 0.5--10\,keV X-ray  spectrum that the  inner disk is
not radially truncated \citep{miller06},  supporting a growing body of
evidence that  disks do not  immediately recede when black  holes move
into the  LHS, and that advective  flows take hold  at lower accretion
rates than is marked by  the state transition.  Finally, \swift\ might  
be placed in the Galactic halo if it is located at a distance of
about 6\,kpc, as has been estimated by Cadolle Bel et al.  (2007) from
its X-ray flux.  The only firmly identified black hole X-ray binaries in
the   Galactic    halo   to   date   are   XTE   J1118+480
\citep{wagner01} and probably BW Cir \citep{casares04}. \\

In  this  paper   we  make  a  detailed  analysis   of  the  long-term
time-resolved  photometry  of  the  optical  counterpart  to  the  BHC
\swift\,  updating  the preliminary  results  presented  in Zurita  et
al. (2007).  The paper is  structured as follows: In $\S$2 we describe
the observations and the data reduction procedure. In $\S$3 we present
the light curves and study  their short and long-term variability. The
results are discussed in $\S$4  where we examine questions such as the
origin of  the periodic modulation  in the light curves,  the distance
towards \swift\ and the possible halo nature of the source.

\section{Observations and data reductions}

\begin{figure}
\includegraphics[angle=0,scale=0.8]{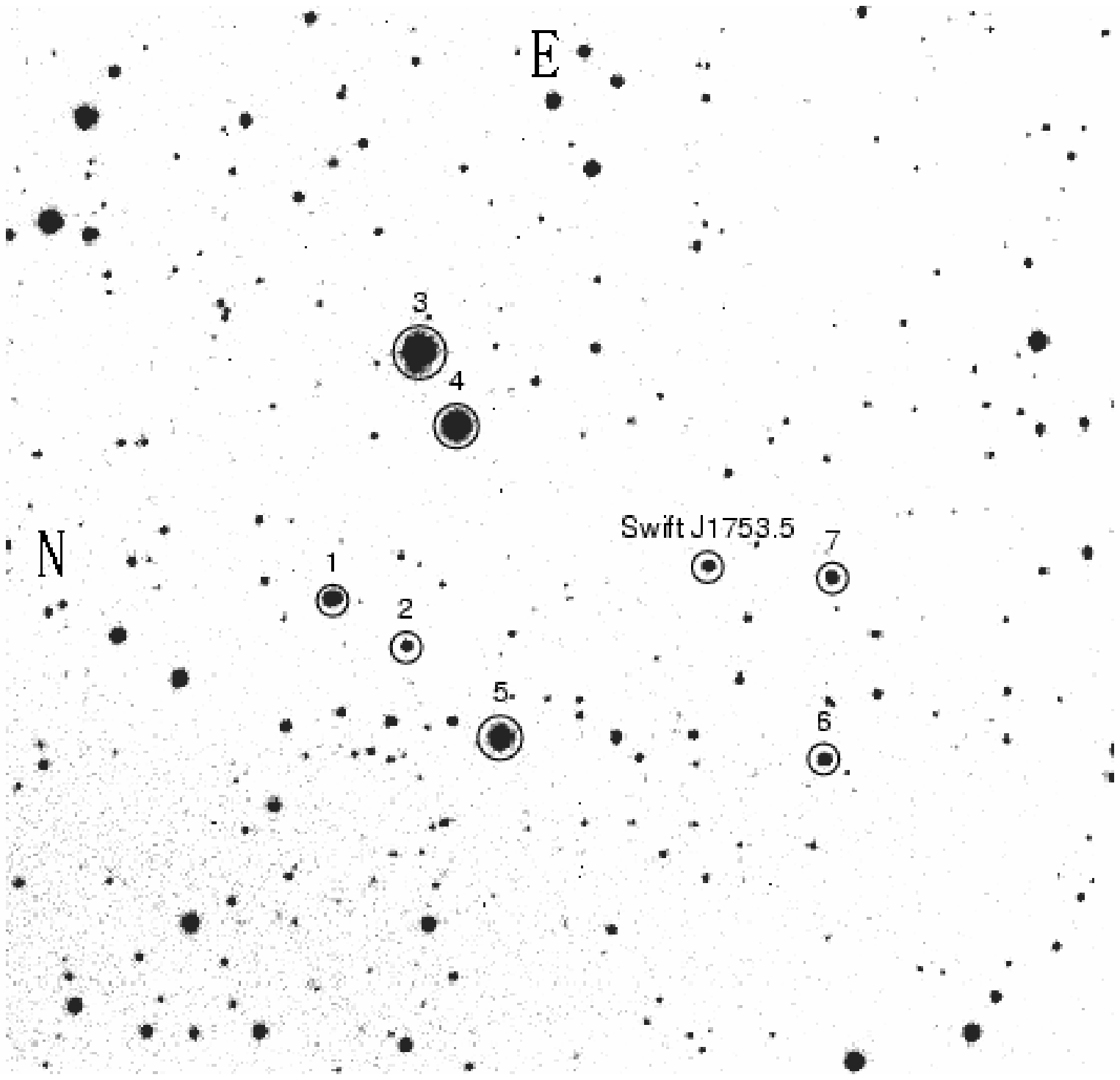}
\caption{  \bf  \label{campo}  R-band 60-s image of \swift. The field of view is 4.2$\times$4.2\,arcmin. The magnitudes of stars 1-7 are listed in Table 2.}
\end{figure}

We observed \swift\ for a total of 20 nights between June 3 and August
31  with  three  different   telescopes:  the  1.5-m  and  the  0.84-m
telescopes at  the Mexican Observatorio Astron\'omico  Nacional on San
Pedro M\'artir, and the 80\,cm IAC80 telescope at the Observatorio del
Teide,  Tenerife, Spain.  The  data consist  of time-series  of R-band
photometry acquired during $\sim$6 h  per night.  The exposure time in
all  sets ranged  from 60  to 120\,s  depending on  the  telescope and
atmospheric conditions. An observing log  is presented in Table 1. All
images were corrected  for bias and flat--fielded in  the standard way
using {\sc iraf}.  We performed aperture or profile fitting photometry
on our object  and several nearby comparison stars  which were checked
for  variability   during  each  night   and  over  the   entire  data
set.  Additionally on  2007 Jun  7 we  obtained $BVRI$  images  of our
target and of 6 Landolt standard fields \citep{landolt92} to perform a
color-dependent   calibration  of   a   set  of   stars  in   the field of \swift\ (see Fig.\ref{campo} and Table 2) .

\begin{table}[h]\caption{Log of observations}\begin{tabular}{ c c c  c c}
\hline
Date  & $\Delta$\,T    & Exp. Time (s) /  &  Mean & Telescope\\ 
      & (h)   & no. exposures    & magnitude & \\
\hline
3/6/2007  & 7.05  &  60 /  293 &      16.42    &  1.5m SPM   \\
4/6/2007  & 6.71  &  60 /  275 &      16.41    &  1.5m SPM   \\
5/6/2007  & 5.23  &  60 /  223 &      16.43    &  1.5m SPM   \\
7/6/2007  & 6.13  &  60 / 272  &      16.44    &   1.5m SPM  \\
8/6/2007  & 5.70  &  60 / 245 &   16.47    &   1.5m SPM  \\
12/6/2007  & 6.87  &  60 /  281 &   16.53    &  IAC80   \\
13/6/2007  & 4.72  &  60 /  210 &   16.50    &  IAC80   \\
14/6/2007  & 6.68  &  60 /  300 &   16.50    &  IAC80   \\
18/6/2007  & 6.93  &  60 /  165 &   16.44    &  IAC80   \\
20/6/2007  & 6.57  &  60 /  295 &   16.51    &  IAC80   \\
21/6/2007  & 5.07  &  60 /  227 &   16.50    &  IAC80   \\
27/6/2007  & 5.09  &  90 /  166 &   16.43  &  84cm SPM   \\
28/6/2007  & 6.90  &  90 /  205 &   16.44  &  84cm SPM   \\
29/6/2007  & 7.39  &  90 /  235 &   16.42  &  84cm SPM   \\
1/7/2007  & 7.52  &  90 /  191 &   16.40  &  84cm SPM   \\
31/6/2007  & 6.85  &  90 /  211 &   16.40  &  84cm SPM   \\
27/8/2007  & 4.26  &  60 /  222 &   16.43  &  1.5m SPM \\
28/8/2007  & 3.77  &  60 /  192 &   16.43  &  1.5m SPM \\
30/8/2007  & 4.52  &  120 / 127 &   16.45  &  1.5m SPM \\
31/8/2007  & 3.18  &  60 /  164 &   16.45  &  1.5m SPM \\
\hline
\end{tabular}
\end{table}

\begin{table}[h]\caption{Magnitudes for seven comparisons in the field of \swift. Errors are less than 0.01\,magnitudes}\begin{tabular}{ c c c c c}
\hline
Star  &  $V$ & $B-V$ & $V-R$ & $V-I$ \\ 
\hline
1 & 16.17  & 1.53 &  0.81 & 1.75  \\
2 & 17.27  & 0.95 &  0.54 & 1.25  \\
3 & 12.49  & 0.60 &  0.37 & 0.93  \\
4 & 13.55  & 0.81 &  0.42 & 1.07 \\
5 & 14.68  & 1.23 &  0.67 & 1.55  \\
6 & 16.66  & 0.96 &  0.49 & 1.19  \\
7 & 16.95  & 1.40 &  0.72 & 1.61  \\
\hline
\end{tabular}
\end{table}

\section{The light curves}

\subsection{Long term behaviour}
The X-ray light curve of \swift\ was typical of XRTs in outburst, with a
Fast   Rise  Exponential  Decay   \citep{ramadevi07}.  It   peaked  at
$\sim$210\,mcrab on  2005 July  6 \citep{cadolle07}.  We  observed the
target  about two  years after  its outburst  and, at  this  time, the
source was still bright in the RXTE ASM energy band (1.2--12\,keV) with
an average flux of $\sim$23\,mcrab.  \\

Immediately after outburst the optical brightness decayed at a rate of
$\sim$0.1\,mag per  week \citep{torres05b}.  The  system had $R=16.15$
on  2005  Aug  11,  $R=16.45$  on  Aug 17  and  $R=16.60$  on  Aug  27
\citep{cadolle07}.  At the  time of  our observations  \swift\ remains
stable   at  $R=16.45$   with  a   night  to   night   variability  of
$\sim$0.05\,mag. The long term light curve and its periodogram are shown in 
Fig.\ref{curvalong}.  The magnitude of
our  object on  the  night  of 2007  June  7 were:  $B=17.00\pm0.005$,
$V=16.60\pm0.005$, $R=16.45\pm0.005$, $I=15.95\pm0.005$.

\subsection{Short term behaviour}

In  Fig.\ref{curvas} we  present the  R-band light  curves  of \swift\
obtained from  2007 June  3 to  August 31.  The  light curves  are not
sinusoidal, but  exhibit a complex morphology  with remarkable changes
in shape and amplitude. Some nights  they have a sawtooth shape with a
$\sim$0.15\,mag peak-to-peak amplitude  whereas other nights they show
a  peculiar   four-peaked  modulation  (more  clear  in   June  29  in
Fig.\ref{curvas})  or  a  flat shape  with  a  deep  dip (as  in  June
12). They resemble the outburst nightly light curves of the black hole
transients   GRO\,J0422+32  \citep{chevalier95,callanan95},  GS2000+25
\citep{charles91} and  Nova Mus 1991  \citep{bailyn92}, all attributed
to  tidal  perturbation  of the  outer  regions  of  the disk  by  the
companion.  Peculiar  superhump modulations showing  several peaks and
dips  are also  often  detected in  cataclysmic variables  \citep[e.g.
V603 Aqr;][]{patterson93}.\\

As the light curves are  non-sinusoidal, we employed the PDM algorithm
\citep{stellingwerf78}  to analyze  the periodicities  present  in the
data.   We  detrendend the  long-term  variations  by subtracting  the
nightly  means   from  the   individual  curves.   This   removed  the
low-frequency  peaks  from  our  PDM spectrum.   The  periodogram  was
computed  in the  frequency range  0.1  to 30  cycle\,d$^{-1}$ with  a
resolution  of  1$\times$10$^{-3}$ and  20  phase  bins.  The  deepest
minimum  is found  at 7.3976  cycle\,d$^{-1}$ which  corresponds  to a
period of 3.2443$\pm$0.0010\,h  (see Fig.\ref{period}).  The frequency
uncertainty was  derived from  the half width  at half maximum  of the
periodogram peak.  A consistent result  (3.2454 +/- 0.0080\,h) is obtained
when calculating a Lomb Scargle  periodogram. The light curves show a complex 
non-sinusoidal shape which explains why the PDM periodogram shows substantial 
power at  2\,P, P  and P/2. We folded all  our data on this  period selected an
origin  at  T$_0$=HJD 2454249.853  which  corresponds  to the  minimum
observed on 2007 Jun 3 (see Fig.\ref{curvafold}).

\section{Discussion}

\subsection{Causes of the 3.24\,h  modulation}

At  the time  of  our observations  \swift\  had a  mean magnitude  of
$V$=16.60, more than 4\,mag  brighter than in quiescence, assuming the
quiescent  magnitude  is $V\ge$21  as  estimated  from  the USNO  B1.0
catalog \cite{monet03}.  Therefore, it  is expected that the companion
star is contributing  only a tiny fraction of  the luminosity and that
the optical flux is dominated by  the accretion disk.  It rules out an
ellipsoidal  modulation  of  the  secondary  as  responsible  for  the
modulation.    It appears very likely that the 3.24\,h periodicity reflects the orbital
motion of  the underlaying binary, through the  reprocessing of X-rays
by the secondary or  superhump modulations.  We regard the possibility
of the minima  occurring near phase 0 are due  grazing eclipses of the
disk by  the secondary  star as unlikely  because the timing  of these
minima is sometimes altered (see Fig.\ref{curvafold}).  Also note that
no X-ray eclipses have been seen.\\

The modulation  can be due to  X-ray heating of  the companion.  X-ray
irradiation, from the inner region  of the accretion disk, can deposit
energy at  a modest optical depth  in the atmosphere  of the secondary
star which is then thermalized  and emerges as optical and ultraviolet
flux.  This  has been  shown to occur  in a  number of low  mass X-ray
binaries in outburst, such as the short-period neutron star transients
4U\,2129+47     \citep{thorstensen79}      and     XTE     J\,2123-058
\citep{zurita00}.  The  optical modulation  in these two  systems was,
however, much larger  than we see in \swift. The  X-ray heating of the
companion is easiest to detect in the ultraviolet where the modulation
can be  very large.   \swift\ was  observed in UV  two days  after the
outburst  onset  \citep{morgan05}   but  no  temporal  variability  on
10-1000\,s   timescales  above   1-sigma   was  found   \citep{still05}.
Furthermore,  optical  colors do  not  vary  significantly with  phase
(Durant et al.  in preparation), which makes this model unlikely.\\

The shape of the light curves suggests that the 3.24\,h periodicity is
a  superhump modulation.   Superhumps are  periodic variations  in the
luminosity of an  accreting binary system, first discovered  in the SU
UMa  class of  dwarf novae,  with a  period slightly  longer  than the
orbital period.  The most  probable explanation for superhumps is they
are due  to the effect of  tidal stresses on  a precessing, elliptical
accretion disk.  The changing shape  and amplitude of the light curves
can  be explained by  changes in  the shape  or size  of the  disk and
resonance between the  Keplerian orbits and the orbital  motion of the
companion.  Superhumps only occur in extreme mass-ratio systems, which
can  form elliptical  accretion disks  \citep[see eg.][]{whitehurst91}
and they have been seen  in XRTs, both in outburst \citep{odonoghue96}
and near  quiescence \citep{zurita02}.  In  the case of \swift\,  a M2
type, or later,  main sequence star is needed to  fit within the Roche
lobe of a 3.24\,h period orbit.  Giving $M_2\le0.3\,M_{\odot}$ for the
secondary star, the  mass ratio may be $q=M_2/M_1<0.1$  if the primary
is  a  black hole  ($M_1>3\,M_{\odot}$).   With  such  low value,  the
eccentric disk should  be forced to precess by  perturbations from the
secondary.  Furthermore,  superhumps show no changes  in colors redder
than U, as is the case here \cite[eg.][]{haefner79}.  For a superhump,
one can  estimate the orbital  period using the  following expression,
empirically   derived   by   Patterson   et   al.    (2005):   $\Delta
P=0.18\,q+0.29\,q^2$,  where $\Delta  P=(P_{sh}-P_{orb})/P_{orb}$ with
$P_{sh}$  and   $P_{orb}$  the   superhump  and  the   orbital  period
respectively and $q$  the mass ratio.  For \swift\  $\Delta P<2\%$ and
thus $3.18\,h<P_{orb}<3.2443\,h$.   Therefore, this is  the black hole
candidate with the shortest orbital period observed to date \citep[a summary list
of orbital periods in black hole binaries can be found in][]{mcclintock06}.\\

\begin{figure}
\includegraphics[angle=0,scale=0.8]{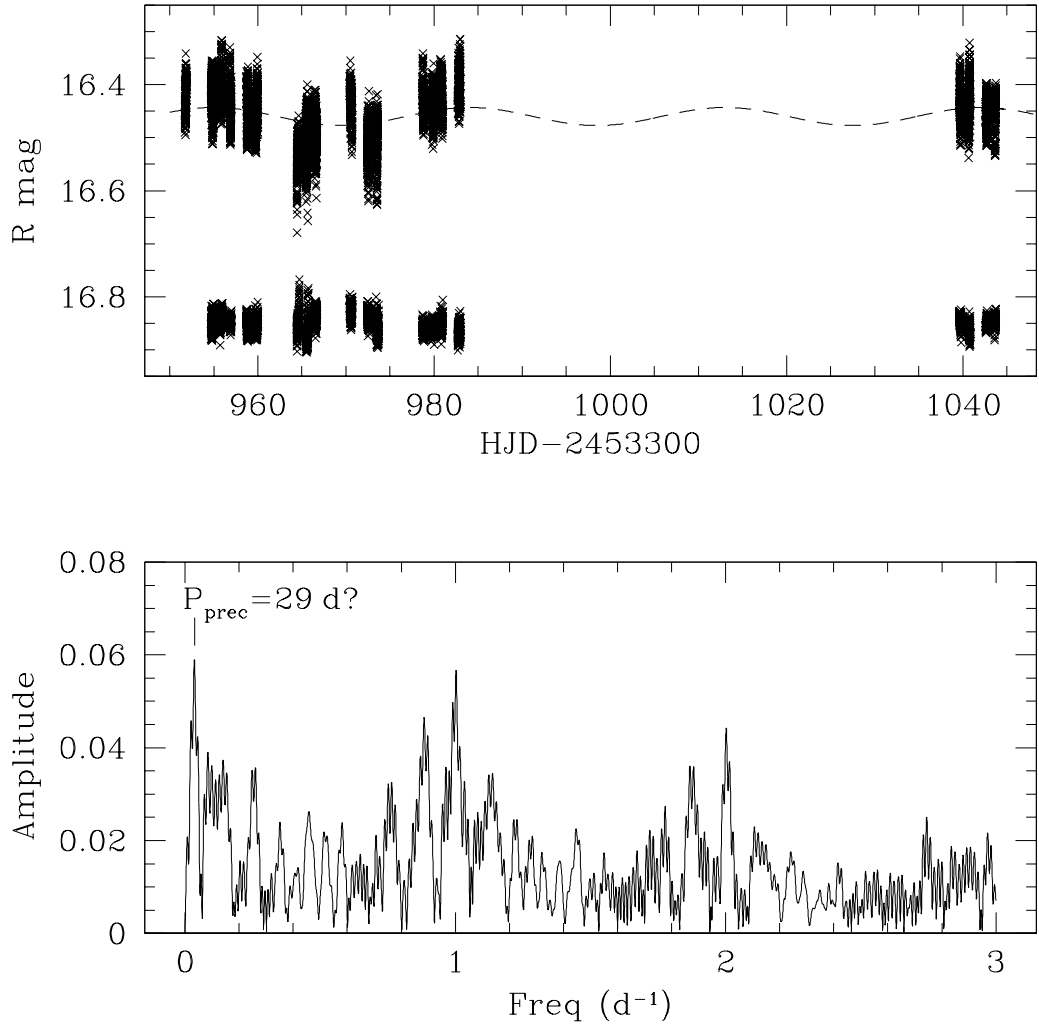}
\caption{  \bf  \label{curvalong}  Long  term R-band  light  curve  of
\swift\  and of  a  comparison  star of  similar  brightness. We  have
overplotted a sinusoidal fit  with 29\,d period (top). Lomb Scargle
periodogram of all our data set from  Jun 3 to August 31 UT. There are
also clear signatures of 1-day alias periodicities.}
\end{figure}

\begin{figure}
\includegraphics[angle=0,scale=0.8]{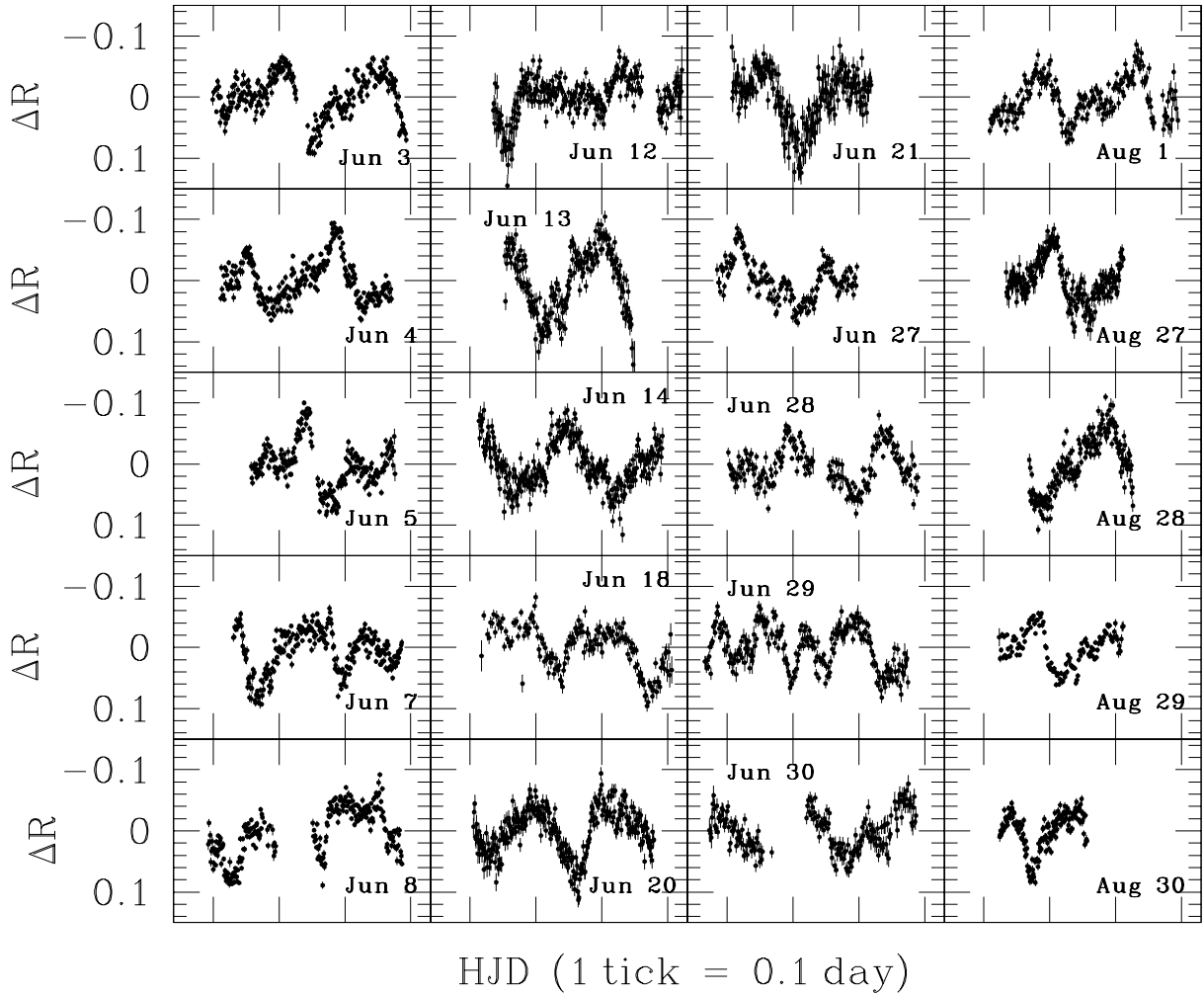}
\caption{ \bf \label{curvas} R-band light curves of SWIFT J1753.5-0127
for the  individual nights. We subtract the  nightly means  from the
individual curves. One tick interval in time is 0.1\,day}
\end{figure}

\begin{figure}
\includegraphics[angle=0,scale=0.8]{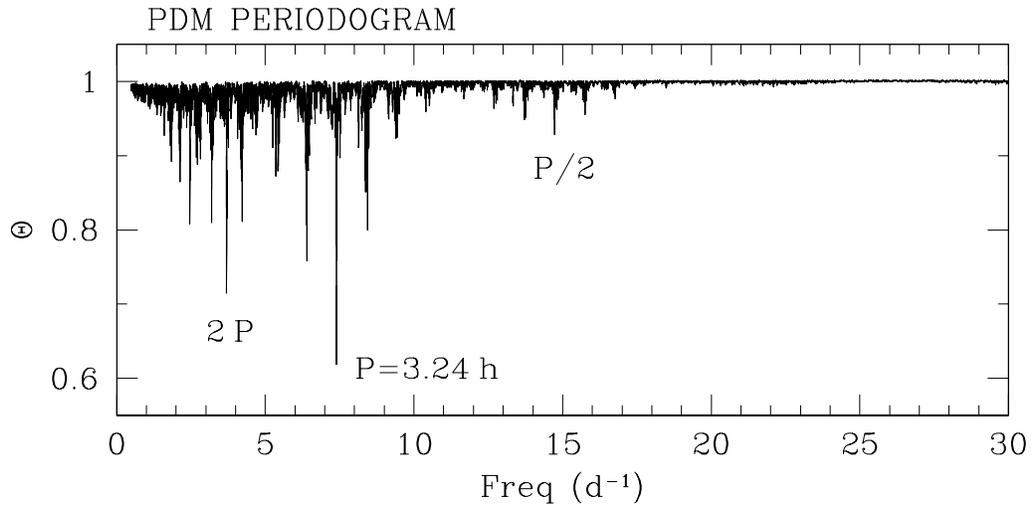}
\caption{  \bf  \label{period} PDM  periodogram  of  the R-band  light
  curves  of all  our  data  set from  Jun  3 to  August  31 UT  after
  detrended the long term  variations by subtracting the nightly means
  from   the   individual  light   curves.    The   deepest  peak   at
  3.24$\pm$0.001\,h and its multiples  2\,P and P/2 are marked.}
\end{figure}

\begin{figure}
\includegraphics[angle=0,scale=0.8]{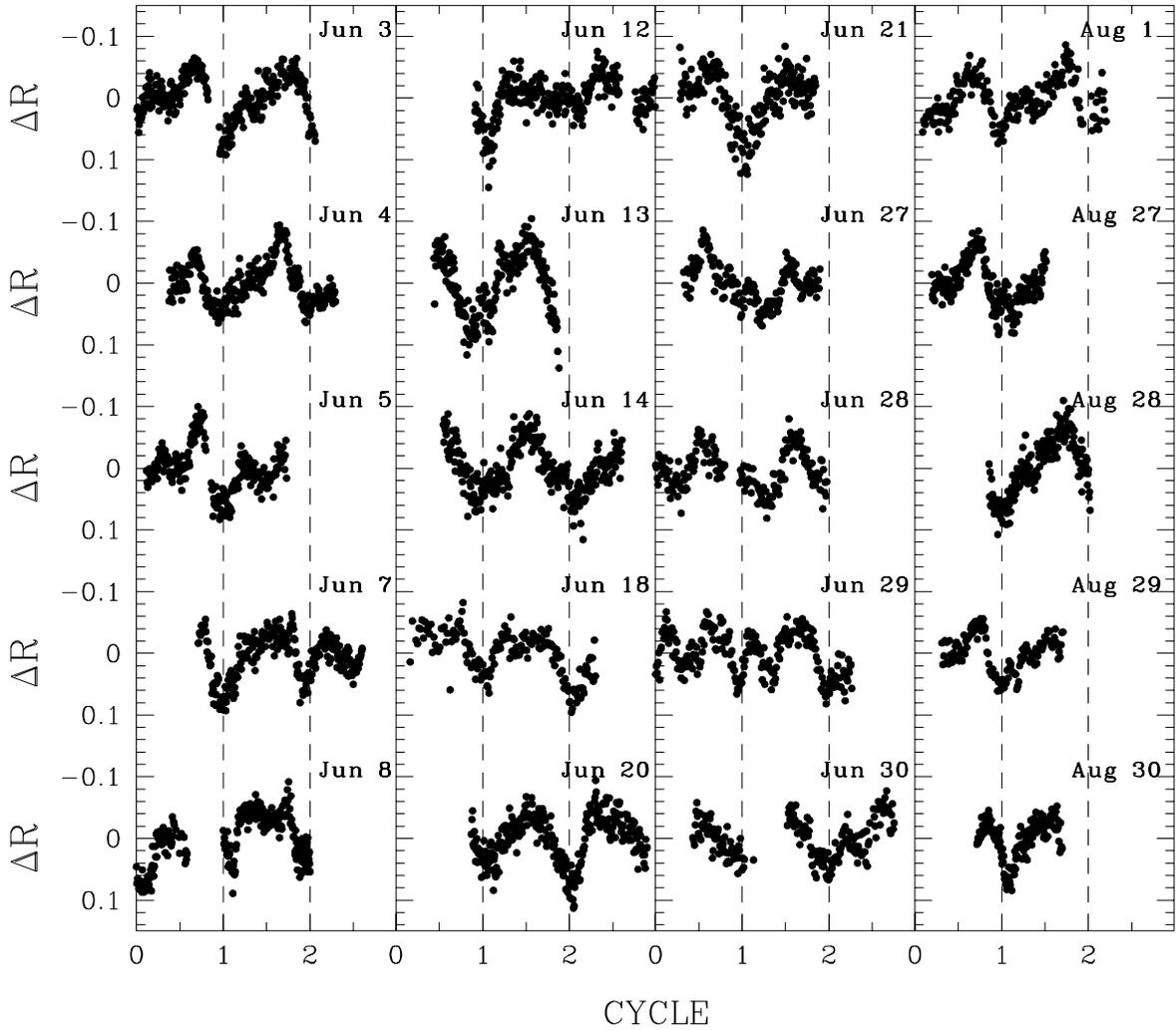}
\caption{   \bf  \label{curvafold}  R-band   light  curves   of  SWIFT
J1753.5-0127  for   the  individual  nights  folded   on  the  3.24\,h
period.  We   selected  an  origin  at   T$_0$=HJD  2454249.853  which
corresponds to the minimum observed on 2007 Jun 3.}
\end{figure}

Since the superhump is the beat frequency between the orbital and disk
precession frequency,  the disk precession period  $P_{prec}$ is given
by   $P_{prec}=(P_{orb}^{-1}-P_{sh}^{-1})^{-1}>7$\,d   (for   a   main
sequence M secondary  and a $>3\,M_{\odot}$ black hole).   In our long
term light  curve, a $\sim$0.1\,mag modulation is  visible (Fig.~1) so
we have  tentatively searched for  periodicities during this  epoch by
calculating a Lomb  Scargle periodogram of our entire  data set in the
frequency range 0 to 3  cycle\,d$^{-1}$.  The highest peak is found at
0.034\,cycle\,d$^{-1}$ which corresponds  to a period of $\sim$29\,day
(Fig.\ref{curvalong}  -- bottom  panel).  This  periodicity,  if real,
could  be associated  to  the  precession of  the  accretion disk.   A
sinusoidal fit with 29\,d period  is shown with the long-term light
curve in  Fig.\ref{curvalong} (top  panel).  Given the  orbital period
implied  by  $P_{prec}$=29\,day,  i.e.,  $P_{orb}$=3.23\,h,  we  infer
$q=0.025$ and  $M_1=12\,M_{\odot}$ for a  M2 type secondary  star. The
confirmation of  the orbital period  and of the secondary  star nature
will  come from radial  velocity variations  and from  observations in
quiescence.

\subsection{A new black hole candidate in the Galactic halo?}

We  can estimate  the  distance $d$  to  the source  by comparing  the
quiescent magnitude  $V_{quiet}$ with the  absolute magnitude of  a M2
type main  sequence star. The quiescent magnitude  is unknown although
$V_{quiet}>$21   according  the   nearby  faint   USNO   B1.0  catalog
\citep{monet03}. With $A_V\,\sim1.05$ \citep{cadolle06} the system has
a distance  $d>1$\,kpc.\\

Futhermore,  Shahbaz  \&   Kuulkers  (1998)  determined  an  empirical
relation for the  absolute magnitude of the accretion  disk during XRT
outburst, from  which we  can estimate the  distances to  the sources.
For systems having  $P_{orb}<$12\,h:

\begin{equation}
5\,log\,d_{kpc}=V_{out}-A_v-2.5\,log\,f-12.34+3.47\,log\,P_{orb}(h)
\end{equation}

where $d_{kpc}$  is the  distance in  kpc and $f$  is the  fraction of
light  arising from  the secondary  star in  quiescence. If  we assume
$f$=1 then the distance estimate is a lower limit. Using this relation
we estimate $d_{kpc}>$7.2 for \swift.\\

Alternatively,  in the  context of  King  \& Ritter  (1998) the  X-ray
exponential  decay  indicates that  irradiation  is  strong enough  to
ionize  the  entire accretion  disk.   Also,  a  secondary maximum  is
expected,  one irradiated-state viscous time  after the  onset  of the
outburst. This can be used to calibrate the peak X-ray luminosity and
hence the distance to the source.

\begin{equation}
d_{kpc}=4.3\times10^{-5}t_s^{3/2}\eta^{1/2}r^{1/2}F_p^{-1/2}\tau_d^{-1/2}
\end{equation}

where $F_p$  is the peak X-ray  flux, $t_s$ the time  of the secondary
maximum after the peak of the outburst in days, $\tau_d$ the e-folding
time of the  decay in days, $\eta$ the  radiation efficiency parameter
and $r$ the ratio of the disk mass at the start of the outburst to the
maximum    possible    mass    \cite{shahbaz98b}.    In    our    case
$F_p$=7.19$\,\times\,$10$^{-9}$\,erg\,s$^{-1}$cm$^{-2}$     in     the
3--25\,keV  energy range,  $\tau_d$=31 \cite{ramadevi07}  and $t_s$=53
from the  X-ray light curve.  Assuming $\eta$=0.05,  and $r$=0.5 we
find $d_{kpc}$=5.6.   For $\eta$  between 0.01--0.05, and  $f$ between
0.5--1 $d_{kpc}$ varies between 2.5 and 7.9\,kpc.\\

The Galactic latitude  of \swift\ is $b^{II}$= 12$^{\circ}$.2  so,
taking a  central value for  the distance of $d$=5.4\,kpc,  the source
should  be at  $z$=1100\,pc  above the  Galactic  plane, rather  large
compared to the distribution of the  BHCs, for which the mean is about
625\,pc \citep{jonker04}.  Among the
BHCs, \swift\ shows  the strongest similarities  with XTE J1118+480  and GRO
J0422+32.  These  latter two  systems have also  been observed  in the
LHS,             exhibit             superhump             modulations
\citep{odonoghue96,uemura00,zurita02}, have  the  shortest orbital
periods:     4.1\,h    \citep{mcclintock01,wagner01}     and    5.1\,h
\citep{filippenko95}  respectively and have  also high   
Galactic  latitudes. It is  hence interesting to  consider  whether \swift\
belongs  to a  population  of Galactic   high-$z$.   The
discovery of  such a  population of halo BHCs in the  Galaxy could
impose new  constraints on  the formation and  evolution of  the black
hole XRTs.

\section{Acknowledgments}

CZ  is  grateful to  Jos\'e  Manuel  Murillo, Antol\'{\i}n  C\'ordova,
Benjam\'{\i}n Garc\'{\i}a  and, in  general, to the  staff of  the San
Pedro M\'artir  Observatory, Baja California,  Mexico. MD acknowledges
support  from the  Spanish Ministry  of Science  and  Technology. MAPT
acknowledges partial support by Chandra grant GO5-6029X. JC akcnowledges support from the MEC through project AYA2006-10254.  
Partially funded by the Spanish MEC under the Consolider-Ingenio 2010  
Program grant CSD2006-00070: First Science with the GTC  
(http://www.iac.es/consolider-ingenio-gtc/)

\end{document}